\documentclass[aps,preprint,amsfonts,amssymb,amsmath,%
tightenlines,nofootinbib,showpacs,showkeys]{revtex4}

\newcommand{\bmu}{\bar{\mu}}
\newcommand{\cN}{\mathcal{N}}
\newcommand{\ee}{\mathrm{e}}
\newcommand{\dd}{\mathrm{d}}
\newcommand{\equal}{\phantom{=\ }}
\newcommand{\AL}{\mathrm{AL}}
\newcommand{\ST}{\mathrm{ST}}
\newcommand{\NP}{\mathrm{NP}}
\newcommand{\PP}{\mathrm{P}}
\newcommand{\re}{\operatorname{Re}}
\newcommand{\im}{\operatorname{Im}}

\begin{document}


\title{On Disagreement about Nonperturbative Corrections in Triple-well
 Potential}


\author{Masatoshi Sato}
\email{msato@issp.u-tokyo.ac.jp}
\affiliation{The Institute for Solid State Physics, The University of Tokyo,\\
 Kashiwanoha 5--1--5, Kashiwa-shi, Chiba 277--8581, Japan}

\author{Toshiaki Tanaka}
\email{ttanaka@ciruelo.fis.ucm.es}
\affiliation{Departamento de F\'{\i}sica Te\'orica II,
  Facultad de Ciencias F\'{\i}sicas,\\
  Universidad Complutense, 28040 Madrid, Spain}


\begin{abstract}

We examine in detail nonperturbative corrections for low lying energies
of a symmetric triple-well potential with non-equivalent vacua,
for which there have been disagreement about asymptotic formulas and
controversy over the validity of the dilute gas approximation.
We carry out investigations from various points of view,
including not only a numerical comparison of the nonperturbative corrections
with the exact values but also the prediction on the large order behavior
of the perturbation series, consistency with the perturbative corrections,
and comparison with the WKB approximation. We show that all the results
support our formulas previously obtained from the valley method calculation
beyond the dilute gas approximation.

\end{abstract}





\maketitle

\section{Introduction}
\label{sec:intro}

It has been widely known that the spectral splitting of the lowest two states
of a quantum mechanical symmetric double-well potential due to the quantum
tunneling is successfully calculated by summing up multiple instanton
contributions with the dilute gas approximation~\cite{Co85}. On the other
hand, it might not have been duly recognized that a naive application of
the method to a bit more complicated system in general confronts some new
difficulties and hardly yields proper results.
In this respect, the problem of a symmetric triple-well potential described
essentially by the following function:
\begin{align}
V(q)=\frac{1}{2}q^{2}(q^{2}-1)^{2},
\end{align}
have been recently attracting attention of several research groups.
One of the novel features of the above potential over the symmetric
double-well potential, besides the obvious difference in the number of
the potential wells, comes from the fact that the harmonic frequency of
the central well is different from that of the side wells. As a result,
there are no degeneracies between the harmonic oscillator spectrum of
the central well and the side wells while that of the left and right well
completely degenerate with each other. Thus, it is difficult to expect
intuitively how the quantum tunneling effect contributes to the each
harmonic oscillator spectrum.

To the best of our knowledge, the multi-instanton calculation technique with
the aid of the dilute gas approximation was applied to a triple-well potential
problem first by Lee \emph{et al.} \cite{LKYPLPY97} and later independently
by Casahorr\'an \cite{Ca01}. Both of their resulting formulas of the lowest
three eigenvalues are however peculiar and doubtful in the fact that they do
not coincide with the harmonic oscillator spectrum of the each potential well
when the instanton contribution due to the quantum tunneling effect is turned
off, although both of the authors did not discuss the validity of them nor
compare the obtained results with the exact eigenvalues.

A few years ago, we investigated in Ref.~\cite{ST02} a similar problem
in a different context, namely, dynamical breaking of $\cN$-fold supersymmetry,
by means of the valley method \cite{AKOSW99}, which is a generalization of the
semi-classical approximation and enables us to calculate nonperturbative
correction beyond the dilute gas approximation. Our formulas of the spectrum
are completely different from the ones previously obtained in
Refs.~\cite{LKYPLPY97,Ca01} and we have justified our results by checking
consistency with some characteristic features of $\cN$-fold supersymmetry
discussed earlier in Ref.~\cite{AST01b}. We have found that the contribution
from the interaction between the instantons plays a crucial role in
the calculation in order to yield the correct formulas. Hence, we have
asserted that the dilute gas approximation would fail in the case.

Recently, however, Alhendi and Lashin reexamined the triple-well potential
problem and carried out a careful calculation of the multiple instanton
contribution with the dilute gas approximation \cite{AL04}. Their results
are different from ours but look sensible in the sense that they reduce to
the harmonic oscillator spectrum when the instanton corrections tend to zero,
in contrast to the ones in Refs.~\cite{LKYPLPY97,Ca01}. They also performed
a numerical calculation of the corresponding Schr\"odinger equation
and compared numerically their formulas with the exact values. From
the comparison, they have claimed the correctness of their formulas and
thus the validity of the dilute gas approximation. However,
they did not make a comparison with our results nor examine the accuracy
of them.

In this letter, considering the present situation described above, we would
like to compare the formulas by Alhendi and Lashin in Ref.~\cite{AL04} and
ours in Ref.~\cite{ST02} from various points of view. In the next
section, we first make a numerical comparison of the nonperturbative
corrections. In Section~\ref{sec:large}, we examine the large order behavior
of the perturbation series. In Section~\ref{sec:inter}, we take into account
the perturbative corrections to check the accuracy of the formulas for
the ground state. In Section~\ref{sec:wkbca}, we carry out the ordinary WKB
calculation for the Schr\"odinger equation to provide another reference for
the comparison. Finally, we summarize the results in the last section.

\section{Purely Nonperturbative Corrections}
\label{sec:nonpe}

In Ref.~\cite{AL04}, Alhendi and Lashin investigated the following triple-well
potential:
\begin{align}
\label{eq:HamAL}
H^{\AL}(x;\omega)=-\frac{1}{2}\frac{\dd^2}{\dd x^2}
 +\frac{\omega^2}{2}x^2 (x^2-1)^2\,.
\end{align}
They calculated the sum of multiple instanton contributions with the dilute
gas approximation and obtained for the lowest three eigenvalues,
\begin{subequations}
\label{eqs:npcALs}
\begin{align}
\label{eq:AL0}
E^{\AL}_{0}(\omega)&=\omega\biggl(\frac{3}{4}-\frac{1}{4}
 \sqrt{1+\frac{1024}{3\pi}\,\omega\ee^{-\omega/2}}\,\biggr),\\
E^{\AL}_{1}(\omega)&=\omega\,,\\
E^{\AL}_{2}(\omega)&=\omega\biggl(\frac{3}{4}+\frac{1}{4}
 \sqrt{1+\frac{1024}{3\pi}\,\omega\ee^{-\omega/2}}\,\biggr).
\end{align}
\end{subequations}
On the other hand, the triple-well potential we investigated in
Ref.~\cite{ST02} is the following:
\begin{align}
\label{eq:HamST}
H^{\ST}(q;g,\epsilon)=-\frac{1}{2}\frac{\dd^2}{\dd q^2}
 +\frac{1}{2}q^2 (1-g^2 q^2)^2 +\frac{\epsilon}{2}(1-3g^2 q^2).
\end{align}
Utilizing the valley method, we obtained for \emph{all} the $n_{0}$th
eigenstates localized around the central well and the $n_{\pm}$th eigenstates
with parity $\pm$ localized around the side wells\footnote{Except for the
cases $\epsilon=\pm(2\cN+1)/3$ $(\cN=0,1,2,\ldots)$ where a part of the
harmonic oscillator spectra of the central and side wells degenerates.},
\begin{subequations}
\label{eqs:npcSTs1}
\begin{align}
E^{\ST}_{n_{0}}(g,\epsilon)&=n_{0}+\frac{1}{2}+\frac{\epsilon}{2}
 +\frac{\sqrt{2}}{\pi g^2}\,\ee^{-1/2g^2}E^{(2)}_{n_{0}}(g,\epsilon)
 +O(\ee^{-1/g^2}),\\
E^{\ST}_{n_{\pm}}(g,\epsilon)&=2n_{\pm}+1-\epsilon
 +\frac{\sqrt{2}}{\pi g^2}\,\ee^{-1/2g^2}E^{(2)}_{n_{\pm}}(g,\epsilon)
 +O(\ee^{-1/g^2}),
\end{align}
\end{subequations}
where the coefficients $E^{(2)}_{n_{0}}$ and $E^{(2)}_{n_{\pm}}$ are given by
\begin{subequations}
\label{eqs:npcSTs2}
\begin{align}
E^{(2)}_{n_{0}}(g,\epsilon)&=-\frac{1}{n_{0}!}\biggl(\frac{2}{g^2}
 \biggr)^{n_{0}}\biggl(-\frac{1}{g^2}\biggr)^{n_{0}/2-1/4+3\epsilon/4}
 \Gamma\biggl(-\frac{n_{0}}{2}+\frac{1}{4}-\frac{3}{4}\epsilon\biggr),\\
E^{(2)}_{n_{\pm}}(g,\epsilon)&=-\frac{(-1)^{(1-3\epsilon)/2}\pm 1}{
 n_{\pm}!}\biggl(\frac{2}{g^2}\biggr)^{2n_{\pm}+1/2-3\epsilon/2}\biggl(
 \frac{1}{g^2}\biggr)^{n_{\pm}}\Gamma\biggl(-2n_{\pm}-\frac{1}{2}+\frac{3}{2}
 \epsilon\biggr).
\end{align}
\end{subequations}
In order to compare the two results, we must first establish the relation
between the Hamiltonians Eqs.~\eqref{eq:HamAL} and \eqref{eq:HamST}. By
applying a scale transformation on the coordinate $q$ in Eq.~\eqref{eq:HamST},
we easily find the following relation:
\begin{align}
\label{eq:scale}
H^{\AL}(x;\omega)=\omega H^{\ST}(\omega^{1/2}x;\omega^{-1/2},0)\,.
\end{align}
Therefore, the quantities which we shall make comparison with
Eqs.~\eqref{eqs:npcALs} are given by\footnote{In this case, the spectral
splitting takes place between the first and second excited states and hence
the parity odd state is lower than the parity even state in the spectrum,
which is in contrast to the case of symmetric double-well potentials.}
\begin{subequations}
\label{eqs:npcSTs3}
\begin{align}
\label{eq:ST0}
E^{\ST}_{0}(\omega)&=\omega\re E^{\ST}_{n_{0}=0}(\omega^{-1/2},0)
 =\omega\biggl(\frac{1}{2}-\frac{\Gamma(1/4)}{\pi}\,\omega^{3/4}
 \ee^{-\omega/2}+O(\ee^{-\omega})\biggr),\\
E^{\ST}_{1}(\omega)&=\omega\re E^{\ST}_{n_{-}=0}(\omega^{-1/2},0)
 =\omega\biggl(1-\frac{4}{\sqrt{\pi}}\,\omega^{3/2}\ee^{-\omega/2}
 +O(\ee^{-\omega})\biggr),\\
E^{\ST}_{2}(\omega)&=\omega\re E^{\ST}_{n_{+}=0}(\omega^{-1/2},0)
 =\omega\biggl(1+\frac{4}{\sqrt{\pi}}\,\omega^{3/2}\ee^{-\omega/2}
 +O(\ee^{-\omega})\biggr).
\end{align}
\end{subequations}
Here we note that the nonperturbative corrections in Eqs.~\eqref{eqs:npcSTs2}
are in general complex and the real parts of them should be taken into account
as the spectral shifts; the imaginary parts of them are to be canceled
with the imaginary parts of the perturbative corrections arising from the
Borel singularity in the framework of the valley method, see for more details
Refs.~\cite{AKOSW99,ST02}. We will later consider the imaginary parts in
order to investigate the large order behavior of the perturbation
series in the next section.

For the purpose of examining the accuracy of the purely nonperturbative
corrections in Eqs.~\eqref{eqs:npcALs} and \eqref{eqs:npcSTs3}, it is
important to note that in addition to the nonperturbative corrections
estimated in the formulas there are perturbative corrections to the
harmonic oscillator spectra\footnote{In this article, a contribution
is called \emph{perturbative} (\emph{nonperturbative}) if it can (cannot)
be expressed as a (formal) power series in $g=\omega^{-1/2}$, respectively.
See also Section~\ref{sec:inter}.}. Hence it does not make sense to compare
directly them to the exact eigenvalues without taking into account
the perturbative contributions. Fortunately, however, the perturbative
corrections to the harmonic oscillator spectrum for the first and second
excited states are completely the same (cf. Section~\ref{sec:large}).
As a result, the difference of the eigenvalues between the first and
second excited states $\Delta E_{21}=E_{2}-E_{1}$ only contains the purely
nonperturbative contributions. Therefore, the comparison of the quantity
$\Delta E_{21}$ enables us to study the accuracy of Eqs.~\eqref{eqs:npcALs}
and \eqref{eqs:npcSTs3} adequately. In Table~\ref{tb:dif12}, we show
i) the exact results $\Delta E_{21}^{ex}$ obtained by solving numerically
the Schr\"odinger equation for the Hamiltonian \eqref{eq:HamAL} presented
in Ref.~\cite{AL04}, ii) $\Delta E_{21}^{\AL}=E_{2}^{\AL}-E_{1}^{\AL}$
obtained from Eq.~\eqref{eqs:npcALs} and the ratios
$\Delta E_{21}^{\AL}/\Delta E_{21}^{ex}$, and iii)
$\Delta E_{21}^{\ST}=E_{2}^{\ST}-E_{1}^{\ST}$ obtained from
Eq.~\eqref{eqs:npcSTs3} and the ratios
$\Delta E_{21}^{\ST}/\Delta E_{21}^{ex}$.\\

\begin{table}[h]
\caption{Comparison of the energy difference between the first and second
excited states.}
\label{tb:dif12}
\begin{center}
\[
\arraycolsep = 8pt
\begin{array}{rl|lc|lc}
\hline
\omega & \Delta E_{21}^{ex}(\omega) & \Delta E_{21}^{\AL}(\omega)
& \Delta E_{21}^{\AL}/\Delta E_{21}^{ex} & \Delta E_{21}^{\ST}(\omega)
& \Delta E_{21}^{\ST}/\Delta E_{21}^{ex}\\
\hline
 30 & 4.7230\cdot 10^{-3} & 3.7381\cdot 10^{-3} & 0.79147
 & 6.8061\cdot 10^{-3} & 1.4411\\
 50 & 9.1006\cdot 10^{-7} & 4.7154\cdot 10^{-7} & 0.51814
 & 1.1081\cdot 10^{-6} & 1.2176\\
 70 & 1.0186\cdot 10^{-10} & 4.1959\cdot 10^{-11} & 0.41193
 & 1.1667\cdot 10^{-10} & 1.1454\\
 90 & 8.9504\cdot 10^{-15} & 3.1490\cdot 10^{-15} & 0.35183
 & 9.9282\cdot 10^{-15} & 1.1092\\
110 & 6.8449\cdot 10^{-19} & 2.1356\cdot 10^{-19} & 0.31201
 & 7.4439\cdot 10^{-19} & 1.0875\\
\hline
\end{array}
\]
\end{center}
\end{table}

{}From Table I, we see that our results $\Delta E^{\ST}_{21}(\omega)$
are in better agreement with the numerical exact values
$\Delta E_{21}^{ex}(\omega)$ than those by Alhendi and Lashin
$\Delta E^{\AL}_{21}(\omega)$. We also find that as the parameter
$\omega$ becomes larger (the coupling g becomes smaller), the accuracy
of our results becomes better while that of the results by Alhendi and
Lashin becomes worse. 
This suggests that our method provides a reliable semi-classical
approximation but the dilute gas approximation does not.
As is easily seen from the Euclidean action of the present model
\begin{align}
S^{\ST}=\frac{1}{g^{2}}\int\dd\tau
 \left(\frac{1}{2}\dot{x}^{2}+\frac{1}{2}x^{2}(1-x^{2})^{2}\right),
 \quad x=gq,
\end{align}
the coupling constant $g^{2}$ plays a similar role to the Plank constant
$\hbar$. Therefore, any reliable semi-classical approximation should work
better as the coupling constant $g$ becomes smaller.

\section{Large Order Behavior of Perturbation Series}
\label{sec:large}

In spite of the fact that perturbation series are in general divergent
and at most asymptotic, they contain much information on the property
of the physical quantity under consideration. An intimate relation
between nonperturbative property and large order behavior of perturbation
series is a typical example~\cite{GZJ90}. In this section, we make
an examination from this point of view.

In order to evaluate the perturbative corrections to the harmonic oscillator
spectrum of the each potential well, it is convenient to begin with
the Hamiltonian \eqref{eq:HamST} and then make the transformation indicated
by the scale relation \eqref{eq:scale} after the perturbative calculation.
The coefficients of the perturbation series are systematically calculated
with the aid of the Bender--Wu method~\cite{BW69}.
First, the perturbation theory around the harmonic oscillator states of
the central potential well is set up by decomposing the Hamiltonian into
the harmonic oscillator part and the remaining part:
\begin{align}
\label{eq:HamST0}
H^{\ST}(q;g,0)=-\frac{1}{2}\frac{\dd^2}{\dd q^2}+\frac{1}{2}q^{2}
 -g^{2}q^{4}+\frac{1}{2}g^{4}q^{6}.
\end{align}
The perturbative corrections to the eigenvalues and eigenfunctions are
defined by the following formal series expansions:
\begin{align}
E(g)=\sum_{m=0}^{\infty}g^{2m}c^{[2m]},\qquad
 \psi(q;g)=\ee^{-q^{2}/2}\sum_{k=0}^{\infty}g^{2k}
 \sum_{l=0}^{\infty}a_{2l+P}^{[2k]}q^{2l+P},
\end{align}
where $P=0$ ($1$) for the even (odd) parity states, respectively.
For the lowest state, $c^{[0]}=1/2$ and $a_{2l}^{[0]}=0$ for all $l>0$.
Requiring that they satisfy the Schr\"odinger equation, we obtain a recursion
relation for $a_{2l+P}^{[2k]}$ and $c^{[2m]}$:
\begin{multline}
\label{eq:recur1}
\qquad (4l+1+2P)a_{2l+P}^{[2k]}-(2l+2+P)(2l+1+P)a_{2(l+1)+P}^{[2k]}\\
 -2a_{2(l-2)+P}^{[2(k-1)]}+a_{2(l-3)+P}^{[2(k-2)]}
 =2\sum_{m=0}^{k}c^{[2m]}a_{2l+P}^{[2(k-m)]}.\qquad
\end{multline}
Second, the perturbation theory around the harmonic oscillator
states of the side potential wells is defined by shifting the origin of the
coordinate to one of the minimum of the side potentials $q\to q\pm 1/g$ and
then decomposing the Hamiltonian:
\begin{align}
H^{\ST}(q\pm 1/g;g,0)=-\frac{1}{2}\frac{\dd^2}{\dd q^2}+\frac{4}{2}q^{2}
 \pm 6g q^{3}+\frac{13}{2}g^{2}q^{4}\pm 3g^{3}q^{5}+\frac{1}{2}g^{4}q^{6}.
\end{align}
The perturbative corrections are similarly introduced by\footnote{Here
we note that all the perturbative coefficients of odd powers in $g$ for
the spectrum vanish due to the parity symmetry of the original Hamiltonian
\eqref{eq:HamST0}.}
\begin{align}
E(g)=\sum_{m=0}^{\infty}g^{2m}c^{[2m]},\qquad
 \psi(q;g)=\ee^{-q^{2}}\sum_{k=0}^{\infty}g^{k}
 \sum_{l=0}^{\infty}a_{l}^{[k]}q^{l}.
\end{align}
The recursion relation for $c^{[m]}$ and $a_{l}^{[k]}$ in this case is then
given by
\begin{multline}
\label{eq:recur2}
\qquad (4l+2)a_{l}^{[k]}-(l+2)(l+1)a_{l+2}^{[k]}\pm 12 a_{l-3}^{[k-1]}\\
 +13 a_{l-4}^{[k-2]}\pm 6 a_{l-5}^{[k-3]}+a_{l-6}^{[k-4]}
 =2\sum_{m=0}^{[k/2]}c^{[2m]}a_{l}^{[k-2m]}.\qquad
\end{multline}
For the lowest state, $c^{[0]}=1$ and $a_{l}^{[0]}=0$ for all $l>0$.

On the other hand, as we have mentioned previously, the imaginary parts
of the nonperturbative contributions are to be canceled with those of
the perturbative ones in the framework of the valley method. This leads to
the following dispersion relation~\cite{ST02,AKOSW99}:
\begin{align}
\label{eq:dispe}
c^{[2m]}=-\frac{1}{\pi}\int_{0}^{\infty}\dd g^2
 \frac{\im E_{\NP}(g)}{g^{2m+2}}\,.
\end{align}
This relation enables us to predict the large order behavior of
the perturbation series for the eigenvalues. For the lowest three states
($n_{0}=n_{\pm}=0$) in the present case ($\epsilon=0$), we obtain from
Eqs.~\eqref{eqs:npcSTs1}--\eqref{eqs:npcSTs2} and \eqref{eq:dispe}
\begin{subequations}
\label{eqs:lobps}
\begin{align}
\label{eq:lobps0}
c_{0}^{[2m]}&\sim -\frac{2^{5/4}}{\pi\Gamma(3/4)}\,2^{m}\Gamma\biggl(
 m+\frac{3}{4}\biggr)\equiv\bar{c}_{0}^{[2m]},\\
c_{1(2)}^{[2m]}&\sim -\frac{8\sqrt{2}}{\pi^{3/2}}\,
 2^{m}\Gamma\biggl(m+\frac{3}{2}\biggr)\equiv\bar{c}_{1(2)}^{[2m]}.
\end{align}
\end{subequations}
Therefore, we can check the validity of the results \eqref{eqs:npcSTs1}--%
\eqref{eqs:npcSTs2} by comparing the predicted asymptotic forms
$\bar{c}^{[2m]}$ in Eqs.~\eqref{eqs:lobps} with the exact perturbative
coefficients $c^{[2m]}$ calculated using the recursion relations
\eqref{eq:recur1} and \eqref{eq:recur2}.

\begin{table}[h]
\caption{The ratios of the exact values of the perturbative coefficients
$c^{[2m]}$ to the predicted asymptotic values $\bar{c}^{[2m]}$.}
\label{tb:large}
\begin{center}
\[
\arraycolsep = 5pt
\begin{array}{rll}
\hline
m & c_{0}^{[2m]}/\bar{c}_{0}^{[2m]} & c_{1(2)}^{[2m]}/\bar{c}_{1(2)}^{[2m]}
 \rule{0pt}{15pt}\rule[-7pt]{0pt}{0pt}\\
\hline
 20 & 0.8946472445 & 0.7797002850\\
 40 & 0.9493285320 & 0.8904365552\\
 60 & 0.9665686152 & 0.9268736279\\
 80 & 0.9750492671 & 0.9451085558\\
100 & 0.9800964967 & 0.9560611732\\
120 & 0.9834448543 & 0.9633690036\\
140 & 0.9858286759 & 0.9685922139\\
160 & 0.9876123158 & 0.9725115700\\
180 & 0.9889971178 & 0.9755611660\\
200 & 0.9901034160 & 0.9780016290\\
220 & 0.9910075563 & 0.9799989041\\
240 & 0.9917603143 & 0.9816636733\\
260 & 0.9923967744 & 0.9830725934\\
280 & 0.9929419564 & 0.9842804381\\
300 & 0.9934141831 & 0.9853273870\\
\hline
\end{array}
\]
\end{center}
\end{table}

In Table~\ref{tb:large}, we show the ratios $c^{[2m]}/\bar{c}^{[2m]}$ up to
the order $m=300$. We easily see that the exact values indeed tend to the
predicted asymptotic values for both the ground and excited states and thus
confirm the correctness of our formulas, at least, for their imaginary parts.

\section{Interplay between Nonperturbative and Perturbative Corrections}
\label{sec:inter}

In Section~\ref{sec:nonpe}, we have investigated the nonperturbative
corrections for the excited states and confirmed that our formulas
obtained from the valley method calculation are more accurate than the ones
obtained from the instanton calculation with the dilute gas approximation.
Although the analysis of the large order behavior
in the previous section ensures the correctness of the imaginary parts
of our formulas for both the ground and excited states, it does not
necessarily mean that the real part of our formula for the ground state
is also correct. In order to check the accuracy of the nonperturbative
spectral shift for the ground state we must resort to other means.

As we have mentioned previously, there exist perturbative corrections
to the harmonic oscillator spectrum in addition to nonperturbative ones.
Hence, each of the spectrum is formally expressed as
\begin{align}
\label{eq:decom}
E(g)=E^{(0)}+E_{\NP}(g)+E_{\PP}(g),
\end{align}
where $E^{(0)}$ denotes the harmonic oscillator spectrum when $g=0$,
$E_{\NP}$ the purely nonperturbative part which cannot be represented by
a formal power series in $g^{2}=\omega^{-1}$, and $E_{\PP}$ the remaining
perturbative part. The decomposition \eqref{eq:decom} suggests that we can
check the formulas for the nonperturbative corrections by examining the
prediction on the \emph{perturbative} corrections instead. That is, we can
regard each of the following quantity as the prediction of the each formula
on the perturbative contribution to the ground state:
\begin{align}
\label{eq:prepc}
E_{\PP}^{\AL}(\omega)\equiv E_{0}^{ex}(\omega)-E_{0}^{\AL}(\omega),\qquad
 E_{\PP}^{\ST}(\omega)\equiv E_{0}^{ex}(\omega)-E_{0}^{\ST}(\omega),
\end{align}
where $E_{0}^{ex}(\omega)$ is the exact eigenvalue for the ground state,
$E_{0}^{\AL}(\omega)$ and $E_{0}^{\ST}(\omega)$ are respectively given
by Eqs.~\eqref{eq:AL0} and \eqref{eq:ST0}, both of which consist of
the harmonic oscillator eigenvalue and the predicted nonperturbative
contribution.

\begin{table}[h]
\caption{Predicted perturbative contributions to the ground state energy.}
\label{tb:prepc}
\begin{center}
\[
\arraycolsep = 5pt
\begin{array}{rll}
\hline
\omega & E_{\PP}^{\AL}(\omega) & E_{\PP}^{\ST}(\omega)\\
\hline
 30 & -0.8\,18251\cdots & -0.8\,21854\cdots\\
 50 & -0.78839\,65537\cdots & -0.78839\,70101\cdots\\
 70 & -0.7763341456\,10396\cdots & -0.7763341456\,51123\cdots\\
 90 & -0.77005483676111\,02506\cdots & -0.77005483676111\,33127\cdots\\
110 & -0.76619756312233715\,58989\cdots & -0.76619756312233715\,61068\cdots\\
\hline
\end{array}
\]
\end{center}
\end{table}
In Table~\ref{tb:prepc}, we show the numerical values of the predicted
perturbative contributions to the ground state energy for each value of
$\omega$ calculated using Eq.~\eqref{eq:prepc}. For the exact eigenvalues
$E_{0}^{ex}(\omega)$, we have used again the numerical results shown in
Ref.~\cite{AL04}.

To examine the accuracy of these predictions, we shall evaluate the
exact perturbative contribution $E_{\PP}$ by using the perturbation series.
Although the perturbation series is generally divergent, as we have
already observed in the previous section (cf., Eqs.~\eqref{eqs:lobps}),
the asymptotic property of the perturbation series nevertheless ensures
that for a sufficiently small value of the expansion parameter a partial
sum of the first finite terms in the perturbation series gives
an asymptotic value of the perturbative correction\footnote{We note that
the asymptotic property of the perturbation series in general has nothing
to do with the Borel summability. We also note that in the quantum mechanical
systems there exist no renormalon singularities which originate from the IR
and UV divergences in the higher-dimensional quantum field theories.}.
As a consequence, however small the value of the expansion parameter is,
there exists a critical order $m_{c}$ at which the absolute value of
the perturbative correction $|g^{2m}c_{0}^{[2m]}|$ becomes minimum. 
It is apparent that the asymptotic property is lost and replaced by
the divergent one when the order of the perturbation exceeds the critical
order $m_{c}$. Therefore, the asymptotic values of the exact perturbative
corrections we should read from the perturbation series are given by
the first finite partial sums up to at most the critical order $m_{c}$.

In Table~\ref{tb:asymp}, we illustrate the numerical values obtained
from the first finite partial sums of the perturbation series for the ground
state. Here we note that from the scaling relation \eqref{eq:scale}
the perturbative quantity $E_{\PP}(\omega)$
we should take for the Hamiltonian $H^{\AL}(x;\omega)$ reads
\begin{align}
E_{\PP}(g)=\sum_{m=1}^{M}g^{2m}c_{0}^{[2m]}\quad\longmapsto\quad
 E_{\PP}(\omega)=\sum_{m=1}^{M}\omega^{1-m}c_{0}^{[2m]}.
\end{align}
In Table~\ref{tb:asymp}, we show the partial sums up to the order $M$
with $m_{c}-10<M\leq m_{c}$. The critical order $m_{c}$ for all the cases
$\omega=30$, $50$, $70$, $90$, and $110$ are given by $m_{c}=\omega/2$.

\begin{table}[h]
\caption{The first finite partial sums of the perturbative corrections.}
\label{tb:asymp}
\begin{center}
\[
\arraycolsep = 5pt
\begin{array}{rcrcrc}
\hline
\multicolumn{2}{c}{\omega=30} &\multicolumn{2}{c}{\omega=50}
 &\multicolumn{2}{c}{\omega=70}\\
M & \sum_{m=1}^{M}\omega^{1-m}c_{0}^{[2m]}
 & M & \sum_{m=1}^{M}\omega^{1-m}c_{0}^{[2m]}
 & M & \sum_{m=1}^{M}\omega^{1-m}c_{0}^{[2m]}
 \rule[-7pt]{0pt}{0pt}\\
\hline
 6 & -0.8\,21307\cdots & 16 & -0.78839\,69801\cdots
 & 26 & -0.7763341456\,49363\cdots\\
 7 & -0.8\,21522\cdots & 17 & -0.78839\,69881\cdots
 & 27 & -0.7763341456\,49752\cdots\\
 8 & -0.8\,21641\cdots & 18 & -0.78839\,69939\cdots
 & 28 & -0.7763341456\,50061\cdots\\
 9 & -0.8\,21714\cdots & 19 & -0.78839\,69983\cdots
 & 29 & -0.7763341456\,50315\cdots\\
10 & -0.8\,21764\cdots & 20 & -0.78839\,70018\cdots
 & 30 & -0.7763341456\,50532\cdots\\
11 & -0.8\,21800\cdots & 21 & -0.78839\,70047\cdots
 & 31 & -0.7763341456\,50723\cdots\\
12 & -0.8\,21830\cdots & 22 & -0.78839\,70072\cdots
 & 32 & -0.7763341456\,50897\cdots\\
13 & -0.8\,21856\cdots & 23 & -0.78839\,70095\cdots
 & 33 & -0.7763341456\,51060\cdots\\
14 & -0.8\,21880\cdots & 24 & -0.78839\,70118\cdots
 & 34 & -0.7763341456\,51217\cdots\\
m_{c} & -0.8\,21903\cdots & m_{c} & -0.78839\,70140\cdots
 & m_{c} & -0.7763341456\,51373\cdots\\
\hline
\end{array}
\]
\[
\arraycolsep = 5pt
\begin{array}{rcrc}
\hline
\multicolumn{2}{c}{\omega=90} &\multicolumn{2}{c}{\omega=110}\\
M & \sum_{m=1}^{M}\omega^{1-m}c_{0}^{[2m]}
 & M & \sum_{m=1}^{M}\omega^{1-m}c_{0}^{[2m]}
 \rule[-7pt]{0pt}{0pt}\\
\hline
36 & -0.77005483676111\,32120\cdots & 46
 & -0.76619756312233715\,61012\cdots\\
37 & -0.77005483676111\,32318\cdots & 47
 & -0.76619756312233715\,61022\cdots\\
38 & -0.77005483676111\,32485\cdots & 48
 & -0.76619756312233715\,61031\cdots\\
39 & -0.77005483676111\,32628\cdots & 49
 & -0.76619756312233715\,61039\cdots\\
40 & -0.77005483676111\,32755\cdots & 50
 & -0.76619756312233715\,61046\cdots\\
41 & -0.77005483676111\,32871\cdots & 51
 & -0.76619756312233715\,61053\cdots\\
42 & -0.77005483676111\,32978\cdots & 52
 & -0.76619756312233715\,61059\cdots\\
43 & -0.77005483676111\,33080\cdots & 53
 & -0.76619756312233715\,61065\cdots\\
44 & -0.77005483676111\,33179\cdots & 54
 & -0.76619756312233715\,61071\cdots\\
m_{c} & -0.77005483676111\,33277\cdots & m_{c}
 & -0.76619756312233715\,61077\cdots\\
\hline
\end{array}
\]
\end{center}
\end{table}

Comparing the results in Table~\ref{tb:asymp} with the ones in
Table~\ref{tb:prepc}, we easily see that the asymptotic values of the
perturbative corrections for all the cases are in good agreement with
the values $E_{\PP}^{\ST}(\omega)$ predicted by our formula \eqref{eq:ST0}
but apparently deviate from the values $E_{\PP}^{\AL}(\omega)$ predicted
by the Alhendi and Lashin's formula \eqref{eq:AL0}.

We note that the differences between $E_{\PP}^{\ST}(\omega)$ and
$\sum_{m=1}^{M}\omega^{1-m}c_{0}^{[2m]}$ are minimum around $M\sim m_{c}-1$.
It indicates that $\re E_{\PP}(g)$ in the case has the perturbation series
as \emph{strong asymptotic series} \cite{RS78}, that is, there exist
positive real constants $C$ and $\sigma$ so that
\begin{align}
\left|\re E_{\PP}(g)-\sum_{m=1}^{M}g^{2m}c_{0}^{[2m]}\right|
 \leq C\sigma^{M+1}(M+1)!\,\left|g^{2}\right|^{M+1},
\end{align}
for all $M$ and all $g^{2}>0$ near the origin. Indeed, if it is the case,
we have with the aid of Eq.~\eqref{eq:lobps0} and $\sigma=2$,
\begin{align}
\left|\re E_{\PP}(g)-\sum_{m=1}^{M}g^{2m}c_{0}^{[2m]}\right|
 \lesssim C'(M+7/4)^{1/4}\Bigl|g^{2(M+1)}c_{0}^{[2(M+1)]}\Bigr|,
\end{align}
where $C'=C\pi\Gamma(3/4)/2^{5/4}$.
The right hand side is minimum around $M\sim m_{c}-1$ by the definition of
the critical order $m_{c}$, and thus the above fact would be naturally
understood.

Finally, we would like to mention about the fact that in the parameter
region we have examined, $30\leq\omega\leq 110$ or
$0.095\lesssim g\lesssim 0.18$, the following relation is roughly satisfied:
\begin{align}
\label{eq:rough}
\min_{m}\Bigl| g^{2m}c^{[2m]}\Bigr|=\Bigl| g^{2m_{c}}c^{[2m_{c}]}\Bigr|
 \sim\bigl|\re E_{\NP}(g)\bigr|\times 10^{-1}.
\end{align}
Interestingly, we can show that a similar relation is observed generically
as far as the system under consideration has a nonperturbative effect.
Suppose the following conditions are satisfied for smaller values of
the coupling constant $g^{2}$ involved in the system under consideration:
\begin{align}
\label{eq:condi}
\im E_{\NP}(g)\sim C g^{-2(\nu +1)}\ee^{-1/bg^{2}},\qquad
 \re E_{\NP}(g)=A \im E_{\NP}(g),
\end{align}
where $A$, $C$, and $b>0$ are real constants. Then, we can prove the
following intriguing relation for smaller $g^{2}$:
\begin{align}
\label{eq:relat}
\min_{m}\Bigl| g^{2m}c^{[2m]}\Bigr|\sim\sqrt{\frac{2b\:\!\ee}{A^{2}\pi}}\,
 \Bigl| g\re E_{\NP}(g)\Bigr|.
\end{align}
For the proof, we first note that the first condition in
Eq.~\eqref{eq:condi} implies
\begin{align}
c^{[2m]}\sim -\frac{C}{\pi}\,b^{m+\nu+1}\Gamma(m+\nu+1)
 \equiv\bar{c}^{[2m]},
\end{align}
for larger $m$. Next, we define a function $f$ by
\begin{align}
f(\mu;g)\equiv\Bigl| g^{2\mu}\bar{c}^{[2\mu]}\Bigr|.
\end{align}
It is evident that for larger integer $m$ the function $f(m;g)$
well approximates the magnitude of the $m$th-order perturbative correction.
The derivative with respect to $\mu$ reads,
\begin{align}
\frac{\partial}{\partial\mu}f(\mu;g)
 =f(\mu;g)\bigl[\,\ln(bg^{2})+\psi(\mu+\nu+1)\bigr],
\end{align}
where $\psi$ denotes the digamma function. Hence $f(\mu;g)$ takes
minimum value at $\mu=\bmu$, $\bmu$ satisfying
\begin{align}
\ln(bg^{2})+\psi(\bmu+\nu+1)=0.
\end{align}
For smaller value of $g^{2}\ll 1$, we see $\bmu$ becomes larger. Thus,
applying the asymptotic expansion of the digamma function~\cite{GR00}:
\begin{align}
\psi(z)\sim\ln z -\frac{1}{2z}+O(z^{-2}),
\end{align}
we obtain
\begin{align}
\label{eq:asymp}
bg^{2}\sim\frac{1}{\bmu+\nu+1}\exp\biggl(\frac{1}{2(\bmu+\nu+1)}
 \bigl[1+O(\bmu^{-1})\bigr]\biggr).
\end{align}
With the aid of the Stirling formula and Eq.~\eqref{eq:asymp}, we have
\begin{align}
\Gamma(\bmu+\nu+1)\sim\sqrt{2\pi}\,\ee^{1/2}(bg^{2})^{-(\bmu+\nu+1/2)}
 \ee^{-1/bg^{2}}\bigl[1+O(\bmu^{-1})\bigr].
\end{align}
Therefore, the minimum value of $f(\mu,g)$, which would provide a good
approximation to the minimum magnitude of the perturbative correction
at the critical order $m_{c}\sim\bmu$, is estimated as,
\begin{align}
f(\bmu;g)
&=\frac{b^{\nu+1}}{\pi}|C|\,(bg^{2})^{\bmu}\Gamma(\bmu+\nu+1)\notag\\
&\sim\sqrt{\frac{2b\:\!\ee}{\pi}}\,|C|\, g^{-2(\nu+1/2)}\ee^{-1/bg^{2}}
 \bigl[1+O(\bmu^{-1})\bigr]\notag\\
&\sim\sqrt{\frac{2b\:\!\ee}{\pi}}\,\Bigl| A^{-1} g \re E_{\NP}(g)\Bigr|
 \bigl[1+O(\bmu^{-1})\bigr],
\end{align}
and thus we obtain the relation \eqref{eq:relat}. In our case $A=1$ and
$b=2$, and thus the relation \eqref{eq:rough} for $g\sim 0.1$ follows.
As a consequence, we also find that the next-order nonperturbative
corrections of order $O(g^{2}\re E_{\NP})$ becomes negligible in comparison
with the perturbative correction for sufficiently small $g$; from
the relation \eqref{eq:relat} we readily obtain
\begin{align}
O\left(g^{2}\re E_{\NP}\right)\sim
 O\left(g\min_{m}\left|g^{2m}c^{[2m]}\right|\right).
\end{align}
Therefore, the next-order nonperturbative corrections do not affect
the analysis for $g\sim 0.1$ in this section.

\section{WKB Calculation}
\label{sec:wkbca}

So far, we have checked the accuracy of the semi-classical calculations
of the path-integral by comparing with the exact values calculated
numerically. In this section, we make a comparison in a different way.
To this end, we employ another nonperturbative approach to derive
formulas for the same physical quantities.
The method we shall use here is the WKB approximation for the Schr\"odinger
equation. In the following, we shall always consider the leading terms of
the power expansion in $g$ since we are interested in the quantization
condition for the nonperturbative contribution.

Let us consider the more general Hamiltonian \eqref{eq:HamST} for all
$\epsilon g^{2}\ll 1$. The system has parity symmetry and thus it is
sufficient to study the connection condition of the WKB wave function
only on the half-line $q\in (0,\infty)$.
In the vicinity of the minimum of the central potential well, the
Schr\"{o}dinger equation for the Hamiltonian \eqref{eq:HamST}
in the leading order of $g$ is given by
\begin{align}
\left(-\frac{1}{2}\frac{\dd^{2}}{\dd q^{2}}+\frac{1}{2}q^{2}\right)\psi(q)
=\left(E-\frac{\epsilon}{2}\right)\psi(q).
\end{align}
The local solutions possessing a definite parity $\pm$ are expressed as
\begin{align}
\label{eq:cent}
\psi(q)=A_{\pm}\left(D_{\nu}(-\sqrt{2}q)\pm D_{\nu}(\sqrt{2}q)\right), 
\end{align}
where $A_{\pm}$ are constants and 
$D_{\nu}$ is the parabolic cylinder function with 
$\nu=E-\epsilon/2-1/2$.
In a similar way, around the minimum of the right side potential well,
the Schr\"{o}dinger equation is approximated by
\begin{align}
\left[-\frac{1}{2}\frac{\dd^{2}}{\dd q^{2}}+2\Bigl(q-\frac{1}{g}\Bigr)^{2}
 \right]\psi(q)=\left(E+\epsilon\right)\psi(q).
\end{align}
and the local solution which vanishes at $q\rightarrow\infty$ is given by
\begin{align}
\label{eq:side}
\psi(q)=B D_{\lambda}\bigl(2(q-1/g)\bigr), 
\end{align}
where $B$ is a constant and $\lambda=E/2+\epsilon/2-1/2$. 
The solutions \eqref{eq:cent} and \eqref{eq:side} are to be connected
with the following WKB solution in the classically forbidden region
($q_1\ll q\ll q_2$):
\begin{align}
\label{eq:WKB}
\psi(q)=\frac{C_{1}}{k(q)^{1/2}}\exp\left(-\int_{q_{1}}^{q}\dd x\,k(x)\right)
+\frac{C_{2}}{k(q)^{1/2}}\exp\left(\int_{q_{1}}^{q}\dd x\,k(x)\right),
\end{align}
where
\begin{align}
k(x)=\sqrt{x^{2}(1-g^{2}x^{2})^{2}+\epsilon(1-3g^{2}x^{2})-2E}\,.
\end{align}
The positive classical turning points $q_{i} (i=1,2)$ with $0<q_{1}<q_{2}$
defined by the solutions of $V(q_{i})=E$ are,
\begin{align}
q_{1}=\sqrt{2E-\epsilon}+O(g^{2}), 
\qquad
q_{2}=\frac{1}{g}-\sqrt{\frac{E+\epsilon}{2}}+O(g).
\end{align}
In order to connect the wave functions obtained in the each region, it
is important to note that the leading term in $g$ of the WKB solution
\eqref{eq:WKB} varies according to the position it is viewed from.
If it is viewed from the point around the central potential well,
the integral in the exponent in Eq.~\eqref{eq:WKB} is evaluated as 
\begin{align}
\int_{q_{1}}^{q}\dd x\,k(x)
&=\frac{1}{g^{2}}\int_{gq_{1}}^{gq}\dd\omega
 \sqrt{w^{2}(1-w^{2})^{2}+\epsilon g^{2}(1-3w^{2})-2E}\notag\\
&=\frac{1}{g^{2}}\int_{gq_{1}}^{gq}\dd\omega\left[w(1-w^{2})
 -\frac{1}{2}\frac{(2E-\epsilon)g^{2}}{w(1-w^{2})}
 -\frac{3\epsilon g^{2}}{2}\frac{w}{1-w^{2}}+\cdots\right]\notag\\
&=\left[\frac{\omega^{2}}{2g^{2}}-\frac{\omega^{4}}{4g^{2}}
 +\frac{\epsilon-2E}{2}\ln|\omega|+\frac{\epsilon+E}{2}\ln|1-\omega^{2}|
 +\cdots\right]_{gq_{1}}^{gq},
\end{align}
and $k(q)\sim q+\cdots$. Thus, in the leading order of $g$ we obtain
the WKB wave function as
\begin{align}
\psi(q)\sim\frac{C_{1}}{q^{1/2}}\,\ee^{-q^{2}/2}\left(
 \frac{\ee\:\! q}{\sqrt{2E-\epsilon}}\right)^{(2E-\epsilon)/2}
\!\!\!\!+\frac{C_{2}}{q^{1/2}}\,\ee^{q^{2}/2}\left(
 \frac{\ee\:\! q}{\sqrt{2E-\epsilon}}\right)^{-(2E-\epsilon)/2}.
\end{align}
Comparing this with the following asymptotic form for $q\gg 1$ of the wave
function \eqref{eq:cent} (cf. Ref.~\cite{GR00}) determined in the region of
the central potential well:
\begin{align}
\psi(q)\sim A_{\pm}\left[\bigl((-1)^{\nu}\pm 1\bigr)\ee^{-q^{2}/2}
 (\sqrt{2}q)^{\nu}+\frac{\sqrt{2\pi}}{\Gamma(-\nu)}
 \ee^{q^{2}/2}(\sqrt{2}q)^{-\nu-1}\right],
\end{align}
we have the following connection condition:
\begin{align}
\label{eq:conn1}
\frac{\sqrt{2\pi}}{\bigl((-1)^{\nu}\pm 1\bigr)\Gamma(-\nu)}
 =\frac{C_{2}}{C_{1}}\left(\frac{2\sqrt{E-\epsilon/2}}{\ee}
 \right)^{2E-\epsilon}.
\end{align}
On the other hand, the integral in the exponent in Eq.~\eqref{eq:WKB}
viewed from the point around the right side potential well is evaluated as
\begin{align}
\lefteqn{
\int_{q_{1}}^{q}\dd x\,k(x)
=\frac{1}{g^{2}}\int_{gq_{1}-1}^{gq-1}\dd\omega
 \sqrt{\omega^{2}(1+\omega)^{2}(2+\omega)^{2}-\epsilon g^{2}
 (2+6\omega+3\omega^{2})-2Eg^{2}}
}\notag\\
&=-\frac{1}{g^{2}}\int_{gq_{1}-1}^{gq-1}\dd\omega
 \left[\omega(1+\omega)(2+\omega)-\frac{\epsilon g^{2}(2+6\omega+3\omega^{2})%
 +2Eg^{2}}{2\omega(1+\omega)(2+\omega)}+\cdots\right]\notag\\
&=\left[-\frac{1}{g^{2}}\left(\omega^{2}+\omega^{3}+\frac{\omega^{4}}{4}\right)
 +\frac{\epsilon+E}{2}\ln|\omega(2+\omega)|+\frac{\epsilon-2E}{2}\ln|1+\omega|
 +\cdots\right]_{gq_{1}-1}^{gq-1},
\end{align}
and $k(q)\sim 1/g-q+\cdots$. Thus, in the leading order of $g$ we obtain
the WKB wave function as
\begin{align}
\psi(q)&\sim\frac{C_{1}\,\ee^{-1/4g^{2}}g^{-3E/2}}{(1/g-q)^{1/2}}\,
 \ee^{(q-1/g)^{2}}\bigl(2(1/g-q)\bigr)^{-(E+\epsilon)/2}\left(
 \frac{\ee}{\sqrt{2E-\epsilon}}\right)^{(2E-\epsilon)/2}\notag\\
&\equal+\frac{C_{2}\,\ee^{1/4g^{2}}g^{3E/2}}{(1/g-q)^{1/2}}\,\ee^{-(q-1/g)^{2}}
 \bigl(2(1/g-q)\bigr)^{(E+\epsilon)/2}\left(\frac{\ee}{\sqrt{2E-\epsilon}}
 \right)^{-(2E-\epsilon)/2}.
\end{align}
Matching this with the following asymptotic form for $1/g-q\gg 1$ of
the wave function \eqref{eq:side} determined in the region of the right
side potential well:
\begin{align}
\psi(q)\sim B\left[\ee^{-(q-1/g)^{2}}\bigl(2(q-1/g)\bigr)^{\lambda}
 -\frac{\sqrt{2\pi}(-1)^{\lambda}}{\Gamma(-\lambda)}\ee^{(q-1/g)^{2}}
 \bigl(2(q-1/g)\bigr)^{-\lambda-1}\right],
\end{align}
we get another connection condition as follows:
\begin{align}
\label{eq:conn2}
-\frac{\sqrt{2\pi}}{(-1)^{\lambda+1}\Gamma(-\lambda)}
=\frac{C_1}{C_2}\,\ee^{-1/2g^{2}}g^{-3E}
 \left(\frac{\ee}{\sqrt{2E-\epsilon}}\right)^{2E-\epsilon}.
\end{align}
Therefore, eliminating the coefficient $C_1/C_2$ in Eqs.~\eqref{eq:conn1}
and \eqref{eq:conn2}, we finally obtain the
following quantization condition:
\begin{multline}
\qquad\frac{\sqrt{2}}{\pi g^{2}}\,\ee^{-1/2g^{2}}
 \frac{(-1)^{E-1/2-\epsilon/2}\pm 1}{2}\biggl(\frac{2}{g^2}
 \biggr)^{E-1/2-\epsilon/2}\Gamma\biggl(-E+\frac{1}{2}+\frac{\epsilon}{2}
 \biggr)\\
\times\biggl(-\frac{1}{g^2}\biggr)^{E/2-1/2+\epsilon/2}\Gamma\biggl(
 -\frac{E}{2}+\frac{1}{2}-\frac{\epsilon}{2}\biggr)=1\,.\qquad
\end{multline}
Strikingly, this condition is in complete agreement with the one obtained
previously by us with the valley method, Eq.~(5.19) in Ref.~\cite{ST02},
from which our formulas of the nonperturbative effect
\eqref{eqs:npcSTs1}--\eqref{eqs:npcSTs2} have been derived. We note that
the coincidence is achieved not only for the case $\epsilon=0$ we have
examined in the preceding sections but also for all $\epsilon\neq 0$
where we cannot apply the usual instanton technique since the classical
configuration now becomes a bounce solution.

\section{Summary}
\label{sec:summa}

In this letter, we have extensively examined lowest three energies
of the symmetric triple-well potential with non-equivalent vacua
by comparing the two different sets of the formulas, the one is calculated
by means of an instanton technique with the dilute gas approximation
in Ref.~\cite{AL04} and the other is by the valley method beyond
this approximation scheme in Ref.~\cite{ST02}.

First, we compared numerically both the formulas with the exact
values for the spectral splitting between the first and second
excited states due to the purely nonperturbative effect. We have found
that in contrast to the latter formulas the prediction of the former
formulas get worse as the value of the parameter tends to the region where
the semi-classical methods should work better. Thus,
contrary to folklore that this kind of problem can be handled by
the use of the dilute gas approximation, our careful comparison with
the exact results indicates that the dilute gas approximation is
insufficient to produce the correct asymptotic formulas even if we restrict
ourselves to examining the low lying eigenvalues.

Furthermore, we examined for the ground state both the perturbative and
nonperturbative contributions. We have found that
the asymptotic values of the perturbative corrections calculated from the
perturbation series deviate from the predicted values obtained from
the instanton calculation with the dilute gas approximation but are in
good agreement with the ones obtained from the valley method calculation.

We also checked the accuracy of the imaginary parts of the latter formulas
by testing the prediction on the large order behavior of the perturbation
series. We evaluated the perturbative coefficients up to the order 300 and
compared them with the predicted asymptotic behaviors. We have confirmed
the correctness of the prediction for both the ground and excited states.

Finally, we carried out the WKB calculation in the leading order of the
coupling constant. The resulting WKB quantization condition is in complete
agreement with the one obtained by means of the valley method in
Ref.~\cite{ST02}, from which our formulas have been derived.
In other words, the dilute gas approximation in the path integral formalism
does not correspond to a proper semi-classical approximation. In this
respect, we would like to recall the fact that discrepancy between the
dilute gas approximation and the WKB approximation has been already reported
in Ref.~\cite{RT83} for the wave functions even in the case of the symmetric
double-well potential.

To conclude, all the present analyses entirely support the valley method
calculation and indicate the limitation of the dilute gas approximation
in the present triple-well potential problem. Therefore, we would like to
repeat the assertion in Ref.~\cite{ST02} that the applicability of the
dilute gas approximation would be quite limited.

\begin{acknowledgments}
We would like to thank the organizers of the international conference
``New Frontiers in Quantum Mechanics'' (July 5--8, 2004, Shizuoka
University, Japan) where the present work started.
This work was partially supported by the Grand-in-Aid for
Scientific Research No.14740158 (M.~S.) and by a Spanish Ministry of
Education, Culture and Sports research fellowship (T.~T.).
\end{acknowledgments}


\bibliography{twprefs}

\begin{thebibliography}{12}
\expandafter\ifx\csname natexlab\endcsname\relax\def\natexlab#1{#1}\fi
\expandafter\ifx\csname bibnamefont\endcsname\relax
  \def\bibnamefont#1{#1}\fi
\expandafter\ifx\csname bibfnamefont\endcsname\relax
  \def\bibfnamefont#1{#1}\fi
\expandafter\ifx\csname citenamefont\endcsname\relax
  \def\citenamefont#1{#1}\fi
\expandafter\ifx\csname url\endcsname\relax
  \def\url#1{\texttt{#1}}\fi
\expandafter\ifx\csname urlprefix\endcsname\relax\def\urlprefix{URL }\fi
\providecommand{\bibinfo}[2]{#2}
\providecommand{\eprint}[2][]{\url{#2}}

\bibitem[{\citenamefont{Colemann}(1985)}]{Co85}
\bibinfo{author}{\bibfnamefont{S.}~\bibnamefont{Colemann}},
  \emph{\bibinfo{title}{{A}spects of {S}ymmetry}}
  (\bibinfo{publisher}{Cambridge Univ. Press}, \bibinfo{year}{1985}).

\bibitem[{\citenamefont{Lee et~al.}(1997)\citenamefont{Lee, Kahng, Yoo, Park,
  Lee, Park, and Yim}}]{LKYPLPY97}
\bibinfo{author}{\bibfnamefont{S.-Y.} \bibnamefont{Lee}},
  \bibinfo{author}{\bibfnamefont{J.-R.} \bibnamefont{Kahng}},
  \bibinfo{author}{\bibfnamefont{S.-K.} \bibnamefont{Yoo}},
  \bibinfo{author}{\bibfnamefont{D.~K.} \bibnamefont{Park}},
  \bibinfo{author}{\bibfnamefont{C.~H.} \bibnamefont{Lee}},
  \bibinfo{author}{\bibfnamefont{C.~S.} \bibnamefont{Park}}, \bibnamefont{and}
  \bibinfo{author}{\bibfnamefont{E.-S.} \bibnamefont{Yim}},
  \bibinfo{journal}{Mod. Phys. Lett.} \textbf{\bibinfo{volume}{A12}},
  \bibinfo{pages}{1803} (\bibinfo{year}{1997}), \eprint{quant-ph/9608015}.

\bibitem[{\citenamefont{Casahorr{\'a}n}(2001)}]{Ca01}
\bibinfo{author}{\bibfnamefont{J.}~\bibnamefont{Casahorr{\'a}n}},
  \bibinfo{journal}{Phys. Lett.} \textbf{\bibinfo{volume}{A283}},
  \bibinfo{pages}{285} (\bibinfo{year}{2001}), \eprint{quant-ph/0103010}.

\bibitem[{\citenamefont{Sato and Tanaka}(2002)}]{ST02}
\bibinfo{author}{\bibfnamefont{M.}~\bibnamefont{Sato}} \bibnamefont{and}
  \bibinfo{author}{\bibfnamefont{T.}~\bibnamefont{Tanaka}},
  \bibinfo{journal}{J. Math. Phys.} \textbf{\bibinfo{volume}{43}},
  \bibinfo{pages}{3484} (\bibinfo{year}{2002}), \eprint{hep-th/0109179}.

\bibitem[{\citenamefont{Aoyama et~al.}(1999)\citenamefont{Aoyama, Kikuchi,
  Okouchi, Sato, and Wada}}]{AKOSW99}
\bibinfo{author}{\bibfnamefont{H.}~\bibnamefont{Aoyama}},
  \bibinfo{author}{\bibfnamefont{H.}~\bibnamefont{Kikuchi}},
  \bibinfo{author}{\bibfnamefont{I.}~\bibnamefont{Okouchi}},
  \bibinfo{author}{\bibfnamefont{M.}~\bibnamefont{Sato}}, \bibnamefont{and}
  \bibinfo{author}{\bibfnamefont{S.}~\bibnamefont{Wada}},
  \bibinfo{journal}{Nucl. Phys.} \textbf{\bibinfo{volume}{B553}},
  \bibinfo{pages}{644} (\bibinfo{year}{1999}), \eprint{hep-th/9808034}.

\bibitem[{\citenamefont{Aoyama et~al.}(2001)\citenamefont{Aoyama, Sato, and
  Tanaka}}]{AST01b}
\bibinfo{author}{\bibfnamefont{H.}~\bibnamefont{Aoyama}},
  \bibinfo{author}{\bibfnamefont{M.}~\bibnamefont{Sato}}, \bibnamefont{and}
  \bibinfo{author}{\bibfnamefont{T.}~\bibnamefont{Tanaka}},
  \bibinfo{journal}{Nucl. Phys.} \textbf{\bibinfo{volume}{B619}},
  \bibinfo{pages}{105} (\bibinfo{year}{2001}), \eprint{quant-ph/0106037}.

\bibitem[{\citenamefont{Alhendi and Lashin}(2004)}]{AL04}
\bibinfo{author}{\bibfnamefont{H.~A.} \bibnamefont{Alhendi}} \bibnamefont{and}
  \bibinfo{author}{\bibfnamefont{E.~I.} \bibnamefont{Lashin}},
  \bibinfo{journal}{Mod. Phys. Lett.} \textbf{\bibinfo{volume}{A19}},
  \bibinfo{pages}{2103} (\bibinfo{year}{2004}), \eprint{quant-ph/0406200}.

\bibitem[{\citenamefont{Guillou and Zinn-Justin}(1990)}]{GZJ90}
\bibinfo{editor}{\bibfnamefont{J.~C.~L.} \bibnamefont{Guillou}}
  \bibnamefont{and}
  \bibinfo{editor}{\bibfnamefont{J.}~\bibnamefont{Zinn-Justin}}, eds.,
  \emph{\bibinfo{title}{{L}arge-{O}rder {B}ehavior of {P}erturbation {T}heory}}
  (\bibinfo{publisher}{North Holland}, \bibinfo{year}{1990}).

\bibitem[{\citenamefont{Bender and Wu}(1969)}]{BW69}
\bibinfo{author}{\bibfnamefont{C.~M.} \bibnamefont{Bender}} \bibnamefont{and}
  \bibinfo{author}{\bibfnamefont{T.~T.} \bibnamefont{Wu}},
  \bibinfo{journal}{Phys. Rev.} \textbf{\bibinfo{volume}{184}},
  \bibinfo{pages}{1231} (\bibinfo{year}{1969}).

\bibitem[{\citenamefont{Reed and Simon}(1978)}]{RS78}
\bibinfo{author}{\bibfnamefont{M.}~\bibnamefont{Reed}} \bibnamefont{and}
  \bibinfo{author}{\bibfnamefont{B.}~\bibnamefont{Simon}},
  \emph{\bibinfo{title}{{M}ethods of {M}odern {M}athematical {P}hysics {IV}:
  {A}nalysis of {O}perators}} (\bibinfo{publisher}{Academic Press},
  \bibinfo{address}{New York}, \bibinfo{year}{1978}).

\bibitem[{\citenamefont{Gradshteyn and Ryzhik}(2000)}]{GR00}
\bibinfo{author}{\bibfnamefont{I.~S.} \bibnamefont{Gradshteyn}}
  \bibnamefont{and} \bibinfo{author}{\bibfnamefont{I.~M.}
  \bibnamefont{Ryzhik}}, \emph{\bibinfo{title}{{T}able of {I}ntegrals,
  {S}eries, and {P}roducts}} (\bibinfo{publisher}{Academic Press},
  \bibinfo{address}{San Diego}, \bibinfo{year}{2000}), \bibinfo{edition}{sixth}
  ed.

\bibitem[{\citenamefont{Rossi and Testa}(1983)}]{RT83}
\bibinfo{author}{\bibfnamefont{G.~C.} \bibnamefont{Rossi}} \bibnamefont{and}
  \bibinfo{author}{\bibfnamefont{M.}~\bibnamefont{Testa}},
  \bibinfo{journal}{Ann. Phys.} \textbf{\bibinfo{volume}{148}},
  \bibinfo{pages}{144} (\bibinfo{year}{1983}).

\end{thebibliography}
\bibliographystyle{apsrev}



\end{document}